\documentclass{article}
\usepackage{LaThuileFPSpro}

\newcommand{\be}{\begin{eqnarray}}
\newcommand{\ee}{\end{eqnarray}}
\newcommand{\bi}{\bibitem}

\newcommand{\mplq}{m_{Pl}^2}
\newcommand{\gmn}{g_{\mu\nu}}
\newcommand{\rmn}{R_{\mu\nu}}
\newcommand{\rv}{\rho_{vac}}
\newcommand{\rde}{\rho_{DE}}
\newcommand{\rc}{\rho_{c}}
\newcommand{\tm}{T_{\mu\nu}^{(m)}}

\begin{document}
\title{ 
PROBLEMS OF VACUUM ENERGY AND DARK ENERGY 
  }
\author{
A.D. Dolgov       \\
  {\em ICTP, Strada Costiera 11, 31014 Trieste, Italy,}\\
{ \em INFN, sezione di Ferrara, Via Paradiso, 12 - 44100 Ferrara, Italy,} \\
{\em ITEP, Bol. Cheremushkinskaya 25, Moscow 113259, Russia.
} 
  }
\maketitle

\baselineskip=11.6pt

\begin{abstract}
A simple description of the vacuum energy (cosmological constant) problem 
for non-experts is presented. Basic features of cosmology with non-zero 
vacuum energy are discussed. The astronomical data which indicate that the
universe is filled with an anti-gravitating state of matter are described.
The mechanisms which may lead to cancellation of almost infinite vacuum 
energy down to the astronomically observed value are discussed. The idea
of dynamical adjustment is considered in some more detail.
\end{abstract}
\newpage
\section{Introduction}
During the last decade it has been established through different independent 
pieces of astronomical data that empty space, devoid of the usual matter, 
is anti-gravitating. It creates gravitational {\it repulsion} and gives rise 
to an accelerated cosmological expansion. According to our present-day
understanding, this accelerated expansion can be induced either by
non-zero vacuum energy, $\rho_{vac}$, 
or by some mysterious agent, called dark energy,
which has positive energy density $\rho_{DE}$ and negative and large
by absolute value pressure, $p_{DE} < -\rde/3$. According to the data,
the magnitude of $\rv$ or $\rde$ is quite close to the critical (closure)
cosmological energy density, 
$\rc \approx 5\, {\rm kev/cm}^3\approx 4\cdot 10^{-47}\,{\rm GeV}^4  $.
Vacuum or dark energy makes approximately 70\% of the latter.

One cannot say that this astronomical discovery was a very big
surprise for physicists working on quantum field theory since
a natural outcome of this theory is a non-zero energy of the
ground state, i.e. $\rv\neq 0$. However, the value of $\rv$, according 
to theoretical expectations is $10^{50} - 10^{120}$ times larger than
the magnitude allowed by cosmology. Strange, but the commonly accepted 
philosophy a decade or more ago was the following: if theory predicts 
something which is almost infinitely big, somehow it must be exactly zero. 
The established non-vanishing value of $\rv$ makes this point of view even 
more vulnerable.

At the present day we face the following three striking problems
which are
very important for our understanding of fundamentals of cosmology and 
field theory:
\begin{enumerate}
\item{}
A huge mismatch between theory and observations at the level 100-50 orders 
of magnitude. Different natural and even experimentally established
contributions to vacuum energy lay in this range. What mechanism is 
responsible for almost complete cancellation of vacuum energy?
\item{}
Why vacuum energy, which must stay constant in the course of cosmological
evolution, or dark energy, which should evolve with time quite differently
from the normal matter, have similar magnitude just today, all being
close to the value of the critical energy density, 
$\rho_c \sim m_{Pl}^2 /t^2$?
\item{}
If universe acceleration is induced by something which is different from 
vacuum energy, then what kind of field or object creates the observed
cosmological behavior? Or could it be a modification of gravitational 
interactions at cosmologically large distances?
\end{enumerate}

The gravity of the problem of vacuum energy was first emphasized (in modern
terms) by Zeldovich\cite{zeldovich-68}, who suggested, in particular, that 
symmetry between bosons and fermions might grossly alleviate it
(this work was a cosmological request for supersymmetry three years
before the idea of supersymmetry was put forward\cite{susy}).
The first attempt to find a mechanism of dynamical adjustment of vacuum energy
from a huge magnitude ``predicted'' by quantum field theory down to $\rho_c$
was made in ref.\cite{dolgov82}. In the mid-eighties the puzzle of $\rv$ 
attracted considerable attention and at the beginning of 90th the first two 
review papers\cite{reviews-1} were published. Now, especially after 
astronomical indications that $\rv$ is indeed non-zero, the number of 
works dedicated to the related problems became explosively large and even the
list of new review papers\cite{reviews-2} is almost outside the limits of 
this article.

The paper is organized as follows. In the next section a brief history of
lambda-term or cosmological constant (both are other names for vacuum
energy) is presented. It is followed (sec.~\ref{s-lam-cosm})
by description of cosmology with a non-zero $\Lambda$.
In sec.~\ref{s-astro} the astronomical evidence in favor 
of non-vanishing $\Lambda$ is discussed. 
Section~\ref{s-contrib} is dedicated to theoretical estimates
of different contributions into $\rv$. In sec.~\ref{s-ways}
possible ways to resolve
huge discrepancy between theoretical expectations and observational data
are described. 

\section{A little history\label{s-hist}}

Cosmological constant was introduced into the gravitational theory
in 1918 in Einstein's paper\cite{ae}, where he tried to  cure
what he thought was a shortcoming of General Relativity, namely
an absence of stationary cosmological solutions. Einstein
noticed that equations of General Relativity (GR) would possess
a stationary solution in cosmological situation if one added an extra term 
proportional to the metric tensor $g_{\mu\nu}$ with a {\it constant} 
coefficient $\Lambda$:
\be
R_{\mu\nu} -{1\over 2} g_{\mu\nu} R = 8\pi G_N T_{\mu\nu}^{(m)} +
g_{\mu\nu} \Lambda,
\label{ein-eq}
\ee
where $G_N \equiv 1/m_{Pl}^2$ is the Newton gravitational constant.
Positive $\Lambda$ could counterweight gravitational attraction of the
usual matter, described by the homogeneous and isotropic energy-momentum 
tensor $T_{\mu\nu}^{(m)}$ and would allow for a stationary solution. This 
solution is evidently unstable because $\Lambda$ must be constant (see below), 
while the energy density of matter decreases if the universe expands and rises 
if it contracts.

After the Hubble's discovery of the cosmological expansion, Einstein 
emphatically rejected the idea of cosmological constant describing it 
as the ``biggest blunder'' of his life. On the other hand, LeMaitre 
considered an introduction of the lambda-term to the theory
as one of the greatest achievements in GR. After the premature 
birth at 1918 the cosmological constant was several times considered dead 
(erroneously) but it looks very much alive today. For a long time a great 
majority of astronomers followed the Einstein's point of view because
cosmology did well without lambda-term and the idea of self-gravitating
empty space looked repulsive, but even at this period LeMaitre, De Sitter, 
and later Eddington were active proponents of cosmological constant. This
was considered as a one-parametric freedom of geometrical General Relativity
theory. Nowadays, in modern language lambda-term is understood as the energy 
density of the lowest energy state, i.e. of vacuum.

At the beginning of sixties astronomical observations 
indicated that recently
discovered quasars are accumulated near redshift $z = 2$. To explain
the observed enhanced quasar population near $z=2$ it was suggested that 
cosmological constant was non-zero and the effect of gravity of the
usual matter was compensated by anti-gravity of $\Lambda$ and hence the
expansion slowed down 
near $z=2$. Many physicists were sceptical about this explanation which
happened to be simply the effect of low statistics and at that time
George Gamow wrote\cite{gg} ``lambda rises its nasty head again''.
After that period for a long time, till the end of nineties a majority of
astronomers, as well as particle physicists believed that cosmological
constant is identically zero. However, at the end of the previous century
several different pieces of data have been accumulated which made it very
difficult (if possible at all) to describe cosmological evolution without
lambda-term. These data are discussed in sec.~\ref{s-astro} but first we
briefly describe some features of cosmology with $\Lambda \neq 0$.

\section{Cosmology with non-zero lambda-term \label{s-lam-cosm}}

Equations of motion (\ref{ein-eq}) implies that $\Lambda = const$ in 
frameworks of metric theory of gravity. Indeed, if one applies the
covariant derivative, $D_\mu$, to both sides of this equation 
the r.h.s. vanishes because of covariant conservation of the 
energy-momentum tensor of matter, 
\be
D_\mu T_{\mu\nu}^{(m)} =0.
\label{dt}
\ee 
The energy momentum-tensor is defined
as the functional derivative of the matter part of action,
$T_{\mu\nu}^{(0)} = \delta S^{(m)} /\delta \gmn$, and its conservation 
follows from the condition that the action is scalar with respect to general
coordinate transformation.

The Einstein tensor $G_{\mu\nu}=\rmn -(1/2)\gmn R$ is automatically transverse,
$D_\mu G^\mu_\nu \equiv 0$ and thus $D_\mu (\Lambda g^\mu_\nu) = 0$. 
From this condition follows that
\be
\partial \Lambda /\partial x^\mu = 0
\label{const-lambda}
\ee
because, by construction, covariant derivative of metric tensor identically 
vanishes, $D_\mu g_{\alpha\beta} = 0$. In some works coordinate dependent
lambda-term was discussed. As is seen from these simple arguments such a 
modification of the theory is not innocent and will not be considered 
here.

According to the modern point of view, cosmological constant is equivalent
to the energy-momentum tensor of vacuum, $T^{(vac)}_{\mu\nu} =\gmn \rv $ and
$\Lambda = 8\pi G_N \rv$. Correspondingly equations (\ref{ein-eq}) can
be rewritten as
\be
\rmn - \frac{1}{2}\, \gmn R = 8\pi G_N \left( \tm + \rv \gmn \right)
\label{new-eq}
\ee

In homogeneous and isotropic case the energy-momentum tensor has the 
diagonal form
\be
T_{\mu}^{\nu} = {\rm diag} \left( \rho,-p,-p,-p\right),
\label{hmgn-t}
\ee
where $\rho$ and $p$ are respectively energy and pressure densities.
For the vacuum one $T_{\mu}^{\nu} \sim g_\mu^\nu = \delta_\mu^\nu$ and
\be
T_{\mu}^{\nu\, (vac)} = {\rm diag} \left( \rho,\rho,\rho,\rho\right),
\label{t-vac}
\ee
Thus vacuum pressure density is negative and large; by absolute value
it is equal to vacuum energy density,
\be
p_{vac} = -\rv
\label{p-vac}
\ee

From this equation immediately follows very interesting and important 
property that energy density of vacuum (or vacuum-like state - see below)
remains constant in the course of cosmological expansion. Indeed, 
the law of covariant energy conservation (\ref{dt})
in homogeneous and isotropic space has the form:
\be
\dot \rho = - 3 H \left( \rho + p \right)
\label{dot-rho}
\ee
Hence for $\rho + p = 0$, as is in vacuum case, $\dot \rv = 0$, this
is a special case of eq. (\ref{const-lambda}) 
Due to this remarkable property the whole huge present-day universe could
be created from a macroscopically small piece of (vacuum-like) matter
during an early inflationary stage.

According to one of the the Friedman equations, the Hubble parameter 
$H= \dot a /a$ is expressed through the energy density (in spatially flat
universe) as
\be 
H = \left( \frac{8\pi \rho G_N}{3}\right)^{1/2}.
\label{H}
\ee
Thus $H=const$ and the expansion is exponential, $a(t) \sim \exp ( H t )$.
In the case when the cosmological expansion is determined by the usual matter,
with $\rho \sim 1/t^2$ then $H\sim 1/t$ and the scale factor rises as a power
of time; $a(t) \sim t^{1/2}$ for relativistic matter and $a(t) \sim t^{2/3}$
for non-relativistic matter. 

Another Friedman equation expresses acceleration of the cosmological
expansion through $\rho$ and $p$:
\be
\frac{\ddot a}{a} = - \frac{4\pi G_N}{3}\, \left( \rho + 3p\right)
\label{ddot-a}
\ee  
For the usual matter, either non-relativistic with $p\approx 0$ or for 
relativistic with $p=\rho/3$, the acceleration is negative, as one should 
expect. Normal matter creates gravitational attraction and slows down 
cosmological expansion. For vacuum the situation is opposite: positive
vacuum energy {\it anti-gravitates}, i.e. creates gravitational repulsion.
The same effect could be created by matter with negative energy density
but such objects are possibly
too exotic to exist. In particular, finite objects with
negative $\rho$ would have negative mass and they not only would create
gravitational repulsion but would move in direction opposite to the direction
of the applied force. Though vacuum state with positive $\rv$ is 
anti-gravitating, it is impossible to create an anti-gravitating object 
of finite
size even if inside it is filled with vacuum(-like) energy. It can be
proved that in the frameworks of normal General Relativity the mass of any
finite object is given by an integral from the energy density and if the
latter is positive, the total mass is also positive. To prove this statement
one has to rely on the Gauss theorem and integration by parts which is 
impossible for infinite systems.

One more remark is in order here. In usual cosmology with vanishing lambda-term
the geometry and ultimate fate of the universe has one-to-one correspondence:
open (and flat) universe will expand forever, while closed universe will 
stop expanding and will re-collapse to a hot and dense stage. This follows
from the first Friedman equation with non-zero curvature term (compare to
eq. (\ref{H}) for flat universe):
\be
H^2 = \frac{8\pi \rho_{tot}}{\mplq} + \frac{k}{a^2}
\label{H2}
\ee
where the constant $k$ determines the curvature of the universe; for
$k>0$ the universe is open and for $k<0$ it is closed. The energy 
density of the usual matter decreases in the course of expansion as $1/a^3$
for relativistic matter and $1/a^4$ for non-relativistic matter. Hence
after sufficiently large time the second term in the r.h.s. of eq. (\ref{H2})
would dominate and if $k<0$ the Hubble parameter should become zero (expansion
stops) and change sign. If the energy density decreases slower then $1/a^2$
(in vacuum case $\rv = const$), then the curvature term may be always
negligible and expansion would last forever with any sign of $k$.

The quantity $\rho_{tot}$ which enters eq. (\ref{H2}) is the sum of
all contributions to cosmological energy density, 
$\rho_{tot} = \rho_m + \rv + \rho_{rel} + ...$. Their relative
fraction is given by the dimensionless ratio
\be
\Omega_a = \rho_a / \rho_c,
\label{omega}
\ee 
where $\rho_c = 3H^2 m_{Pl}^2/8\pi$ is
the critical energy density. It is evident that if $k=0$, that is the
universe is geometrically flat, i.e. $\Omega_{tot} = 1$.

Strictly speaking it is not established that the new form of matter/energy
observed by astronomers is indeed vacuum energy. This new contribution
to cosmological energy density got the name dark energy. By assumption
its equation of state can be written in the same way as equation of state
of other forms of matter/energy:
\be
p_{DE} = w \rho_{DE} 
\label{eqn-st}
\ee
As we have mentioned above $w=0$ for non-relativistic matter, $w=1/3$ for
relativistic matter, and $w=-1$ for vacuum, see eq. (\ref{p-vac}). For
an accelerated cosmological expansion one needs $w<-1/3$, as follows from
eq. (\ref{ddot-a}).

At this stage is worthwhile to make a remark about possible matter fields 
which could mimic vacuum energy, but possibly with $w\neq -1$. The simplest
example is given by a slowly varying scalar field $\phi$. Its 
energy-momentum tensor is
\be
T_{\mu\nu} = 2 \phi_\mu \phi_\nu - \gmn \left[ \phi_\alpha \phi^\alpha -
U(\phi) \right]
\label{tmunu}
\ee
where $\phi_\mu = \partial \phi/\partial x^\mu$ and $U(\phi)$ is the
$\phi$-potential. It is easy to see that if derivatives $\phi_\mu$ are
small then the energy-momentum tensor is dominated by the potential 
term $U(\phi)$ and has the vacuum-like form, i.e. it is proportional
to $\gmn$. This could be realized if the potential $U(\phi)$ is
sufficiently smooth. Another possibility is that $U(\phi)$ has 
a local minimum with $U(\phi) \neq 0$. In such a case the system might 
stuck in this minimum (false vacuum) for very long time and the
cosmological constant would be non-zero till false vacuum explodes.
Depending on the order of the phase transition to real vacuum, the
process might be smooth and quiet (second order phase transition) or
really explosive. 

Parameter $w$ for a homogeneous scalar field in homogeneous cosmological 
background is equal to
\be
w (\phi) = -\frac{2U(\phi) - \dot \phi^2}{2U(\phi) + \dot \phi^2}
\label{w-phi}
\ee
From this expression one can easily see that $-1<w<+1$,
if $U(\phi) >0$. In fact this
is true for any normal theory. To have $w<-1$, theory should be quite
pathological, see refs.\cite{dolgov-s2,phan}. If such 
a large negative value
of $w$ is realized cosmological expansion will end up with crushing
singularity. As we see in what follows, the present-day data does not 
exclude $w<-1$, but most probably the simplest possibility $w=-1$ 
(or $w>-1$) is realized.

\section{Observational data \label{s-astro}}

There are several independent pieces of astronomical data which require
non-zero vacuum energy or, similar to it, dark energy.
\begin{enumerate}
\item{}
Direct observations of cosmological acceleration.
\item{}
Measurements of the curvature of the universe.
\item{}
Measurements of the the total cosmological density of the usual matter.
\item{}
Theory and observation of large scale structures in the universe.
\item{}
Data on the universe age.  
\end{enumerate}
Corresponding astronomical tests are discussed in many standard textbooks
on cosmology. For recent reviews one can see e.g. refs.\cite{cosm-rev}.
All relevant observations unanimously require
\be
\Omega_m \approx 0.3\,\,\, {\rm and}\,\,\, \Omega_\lambda \approx 0.7
\label{om-m-om-lam}
\ee
Here $\Omega_\lambda$ is the cosmological fraction either of vacuum or
dark energy. An important feature that strongly amplify the reliability
of this result is that it is obtained not only from measurements of the 
same quantity by different instruments and methods but also from measurements
of different and unrelated effects. For example, acceleration of the 
universe is observed through luminosity of high redshift supernovae,
SNIa. Absolutely independent measurements of angular fluctuations of
CMBR also demand vacuum-like energy to be non-vanishing and of quite
close magnitude. Theory of large scale structure formation strongly
supports non-zero $\Omega_{\lambda}$ too. Moreover, measurements of density
fluctuations from different sides: from large scales by CMBR and from 
small scales by study of matter distribution in the universe (so called
large scale structure) demonstrated perfect agreement in coinciding
range of wave lengths. This shows that the main features of modern
cosmology with non-zero $\Omega_{lambda}$ are basically correct.

Below in this section we will briefly discuss some of 
astronomical data which demands non-zero (and large) $\Omega_\lambda$
and small $\Omega_m<1$.

\subsection{Cosmological acceleration \label{ss-accel}}

To measure cosmological acceleration one needs to measure universe 
expansion rate at large distances or, what is the same, at high 
redshifts $z\simeq 1$. Of course some deviations from a constant
speed expansion do exist at close distance, but they are very small. 
If there are astronomical objects (standard candles) with known luminosity, 
$L$, then their observation would allow to determine the
flux-redshift relation. The measured flux, $f$, permits to determine the 
distance to the object, $d_L = (L /4\pi f)^{1/2}$, while redshift (by
definition) is determined by the Doppler shift of spectral lines. The 
distance to the object can be expressed through redshift, the present day 
value of the Hubble parameter and the law of the evolution of the latter
with time. For decelerated expansion the distance would be shorter then
for the case of constant speed expansion, while for accelerated expansion
the distance would be larger. In the first case the objects would be 
brighter, while in the second they would be dimmer. 

Possibly good standard candles are type Ia supernovae. At least those
observed nearby seem to be such. Observations\cite{accel,tonry-03} of high 
red-shift SN Ia show that they are systematically dimmer than would be
expected for normal decelerated expansion and thus a possible conclusion
could be that the universe expands with acceleration and vacuum energy may
be non-vanishing. Though there already existed several other pieces of 
astronomical data indicating in the same direction, this 
discovery\cite{accel} made the final blow to an old point of view that 
vacuum energy must be identically zero.

Due to great importance of this result one should be very cautious 
and try other possible explanations of supernova dimming. Dust present
in intergalactic medium would suppress the flux from distant object but
simultaneously the usual dust (with the particle size comparable to
the light wave length) should shift the color of the light towards
red. This effect was not observed. 

There could be so called grey dust with particle size much larger 
than the wave length. Such dust would diminish the flux leaving 
spectrum intact. However, the recent data\cite{tonry-03}
show that the supernovae observed
at higher redshift, $z\geq 1$ become brighter. It is exactly
what should be expected if supernova dimming is created by vacuum 
or vacuum-like energy. Indeed, as we have seen above, vacuum energy 
remains constant in the course of cosmological evolution, while
the energy density of non-relativistic matter evolves as $(z+1)^3$.
If the ratio of vacuum to matter energy densities today is 7/3, as is 
found from observations, 
then at redshifts above $z_{eq} = 0.67$ the difference
$2\rv - \rho_m$ becomes negative and the expansion would be ``normal''
decelerated. Thus at $z>z_{eq}$ one would expect that the dimming of 
supernova with respect to the ``constant speed'' 
expansion should gradually
decrease and ultimately turn into brightening. The observation of 
non-monotonic behavior of dimming with redshift makes also unlikely
a possible explanation of dimming by SN evolution effects.

Thus it seems that the most natural explanation of the dimming of
 the high redshift SN Ia is the accelerated expansion of the universe. 
The data are sensitive, roughly speaking, to the difference of 
$\rv$ and $\rho_m$
and does not allow to determine both but together with other
astronomical observations a separate determination of density of the
usual (dark) matter and dark energy is possible. In particular, the
recent analysis\cite{tonry-03} yields 
$\Omega_\lambda -1.4 \Omega_m = 0.35 \pm 0.14$, under assumption that 
the equation of state of dark energy is $w=-1$. The data 
of ref.\cite{tonry-03} are compatible with this assumption 
giving $w = -1.02^{+0.13}_{-0.19}$. If the universe
is flat (see the next subsection) i.e. 
$\Omega_{tot} =\Omega_m+\Omega_\lambda =1$, then 
$\Omega_m = 0.29^{+0.05}_{-0.03}$, independently on the data on the 
large scale structure (LSS) of the universe.

\subsection{Curvature of the universe}

The curvature of the universe can be accurately determined from the
classical cosmological angular size test applied to angular fluctuations 
of cosmic microwave background radiation (CMBR) (for a simple review see
e.g. ref.\cite{cmbr-rev}). The first (highest) acoustic peak in the
angular spectrum of fluctuations corresponds to the sound wave with the
length equal to the so called sound horizon $d_s$ at the epoch of hydrogen
recombination. The latter differ from the Hubble horizon by the speed of
sound factor, $c_s = 1/\sqrt {3}$. If we know $d_s$ and know the angular 
size at which we observe it today we are able to say what 
is the geometry of 
the universe. For an open universe the angle would be bigger that for the
flat one, while for the closed universe the angle would be smaller. All
recent measurements of angular fluctuations of CMBR show 
the position of the
first acoustic peak exactly at the place which corresponds to the flat
universe. According to the most precise data of WMAP\cite{wmap} combined 
with other astronomical data the result is: 
\be
\Omega_{tot} = 1.02 \pm 0.02  
\label{omega-tot}
\ee
Thus in addition to the ``difference'' of $\rv$ and $\rho_m$ their
sum is also measured. From these one can find $\rv$ and $\rho_m$
separately as is described in the previous subsection. Moreover,
the positions and heights of acoustic peaks in CMBR angular 
spectrum depend upon the law of cosmological expansion and hence
upon the fraction of non-relativistic matter. Thus
an analysis based on the complete set of WMAP data\cite{wmap} (and
not only on the position of the first peak) allows to determine
$\Omega_m$ without invoking results of other measurements and
gives $\Omega_m = 0.29\pm 0.07$ in good agreement with direct
determination of the latter. If one includes data and theory of LSS 
formation then the fraction of matter would be 
$\Omega_m = 0.27\pm 0.04$\cite{wmap}. 

Simultaneously one can determine
the equation of state of dark energy. According to ref.\cite{wmap}
it is $w= -0.98\pm 0.12$.

\subsection{Mass density of usual (dark) matter}

There are several ways to determine the
cosmological energy density of 
clustered matter\footnote{According to eq. (\ref{const-lambda}),
vacuum energy must be uniform, however one
should bear in mind that if dark energy is not just
simple vacuum energy, but the energy of some new weakly interacting 
field, then depending upon the properties of this field, dark energy
may also be clustered.}. One possibility is to study the galaxy velocity
fields assuming that galactic peculiar motion (with respect to the Hubble
flow) is induced by gravitational action of surrounding matter. Different
methods and samples of galaxies give consistent results\cite{gal-vel}
averaging around $\Omega_m = 0.3$. 

The study of equilibrium of hot gas in rich clusters permits to measure
the ratio of the mass of the baryonic component to the total mass of the
cluster. This ratio was found to be quite large, 
$\Omega_b/\Omega_m \approx 0.15$. Since it is known from BBN and 
independently from CMBR that $\Omega_b \approx 0.05$, we find
again $\Omega_m \approx 0.3$.

The third method of determination of the mass density of matter in 
the universe is based on determination of cluster abundances at
different redshifts z. In the universe with $\Omega_m \sim 1$
cluster formation strongly grows with time and number of clusters today
must be much larger than, say, at $z\simeq 1 $. On the contrary, in
low $\Omega_m$-universe the number of clusters at the present time
and around $z=1$ should be approximately the same. Observations 
demonstrate very small change in cluster abundances and thus support
low mass cosmology. Analysis made in different works\cite{cluster-z} 
lead to the conclusion that $\Omega_m = 0.1-0.4$. Though
the dispersion of the results is quite high, they reliably exclude
large values of $\Omega_m$. 

Matter inhomogeneities along the line of sight to background galaxy
distort its image due to gravitational lensing effect. This is a basis
of one more method of ``weighting'' the universe. Results of different
measurements are summarized in review\cite{grav-lns} and all agree
with low $\Omega_m$. 

Thus we see that several independent astronomical measurements give 
the consistent result $\Omega_m \approx 0.3$. 

\subsection{Universe age}

Long existing discrepancy between a relatively large value of the Hubble
parameter $H\approx 70$ km/sec/Mps and large universe age, $t_U$,
is nicely resolved if vacuum energy is non-zero. If we compare two
regimes of cosmological expansion, accelerated and decelerated, then
with the same value of the Hubble parameter at the present time
$H=\dot a/a$, expansion was slower in the past for accelerated regime.
it means that to reach the same magnitude of $H$ more time was 
necessary and the accelerated universe should be older. 

The universe age can be expressed through the present day values of
the Hubble parameter $H_0$ and fractions of different forms 
of energy as:
\be
t_u = \frac{1}{H_0} \int_0^1 \frac{dx}{\left(1-\Omega_{tot} +\Omega_m/x
+\Omega_v x^2\right)^{1/2}}  
\label{tu}
\ee
where $H_0^{-1} = 9.8\cdot 10^9 h^{-1}$ years and dimensionless parameter $h$
according to modern data is about 0.7. Hence in flat matter dominated 
universe with $\Omega_{tot} = \Omega_m =1$ the universe would be only
9.3 Gyr while nuclear chronology (reviewed in ref.\cite{nuc-chron}) 
and the age of old
globular clusters (reviewed in ref.\cite{glob-clus}) indicate much larger 
age, $t_u = 12-15$ Gyr. For flat universe with $\Omega_m= 0.3$ and
$\Omega_v = 0.7$ the universe age according to eq. (\ref{tu})
is $t_u = 13.8 $ Gyr in good agreement with the range quoted above.

\section{Contributions to vacuum energy \label{s-contrib}}

Quantum field theory and particle physics predict that there are several
huge contributions into vacuum energy, while we see from observations 
that all these contributions are miraculously canceled almost to nothing
on the scale of particle physics but just of order of unity on the 
present day cosmological scale.

It is well known from quantum mechanics that the ground state energy
of an oscillator is not zero but $E_0=\omega/2$. Quantum field theory
deals with infinitely many oscillators labeled by their wave number
${\bf k}$ and the energy of the ground state (vacuum) is infinitely
large. For example the energy density of a bosonic field is given by
the integral:
\be
\langle {\cal H}_b \rangle_{vac} =g_s \int \frac{d^3 k}{(2\pi)^3}\,
\frac{\omega_k}{2}  
\langle a^\dagger_k a_k + b_k b^\dagger_k \rangle_{vac} =
g_s\int \frac{d^3k}{(2\pi)^3}\,\omega_k = \infty^4
\label{h-b}
\ee 
where $g_s$ is the number of spin states,
$a_k^\dagger$ and $a_k$ are creation-annihilation operators for 
particles, $b_k$ are the same for antiparticles, and 
$\omega = \sqrt{k^2 +m^2}$. 

Vacuum energy of a fermionic field is given by a similar integral
with a sign difference:
\be
\langle {\cal H}_f \rangle_{vac} = g_s\int \frac{d^3 k}{(2\pi)^3}\,
\frac{\omega_k}{2}  
\langle a^\dagger_k a_k - b_k b^\dagger_k \rangle_{vac} =
g_s \int \frac{d^3k}{(2\pi)^3}\,\omega_k = -\infty^4
\label{h-f}
\ee 
Thus if the theory is symmetric with interchange of bosons and fermions
i.e. there exist an equal number of bosonis and fermionic
states with equal masses
(pairwise) then the energy of vacuum fluctuations would naturally be
zero\cite{zeldovich-68}.

Later, when supersymmetry was theoretically discovered\cite{susy} it was 
found that indeed $\langle {\cal H}_{tot} \rangle_{vac} =0$ if the
symmetry were unbroken. We know, however, that supersymmetry is
broken and supersymmetric partners of
the known particles, if they exist, should be much heavier, at least
in 100 GeV-TeV range or even above. Moreover, theoretically preferred
breaking of supersymmetry, so called soft breaking, demands that vacuum 
energy is non-zero:
\be
\langle {\cal H}_{tot} \rangle_{vac} \sim m_{susy}^4 \geq 10^8\,\, {\rm GeV}
\geq 10^{54} \rho_c
\label{h-susy}
\ee 
Hence one should exclude the most appealing theoretically
models of supersymmetry breaking
but there exists a local generalization of global supersymmetry, namely
supergravity which is free from this constraint. Still the natural value
of vacuum energy in supergravity is about 
$m^4_{Pl} \sim 10^{76}$ GeV and its near vanishing demands fine-tuning of
the parameters more then by 120 orders of magnitude.

Theories with spontaneously broken symmetry create an additional problem
for nullification of vacuum energy. Effective potential of the scalar
(Higgs) field responsible for spontaneous symmetry breaking typically has
the following form:
\be
U(\Phi) = - m^2_\Phi \Phi^2 + \lambda \Phi^4 = 
\lambda (\Phi^2 - \eta^2)^2 + m^4_\Phi/\lambda
\label{u-higgs}
\ee
In broken symmetry phase $\Phi = \eta$ and the last term in the above 
equation leads to huge vacuum energy $\rv = m^4_\Phi/\lambda$. We have
to believe that, by some mysterious reason, this term should be subtracted 
from these expressions. In other words, vacuum energy in unbroken phase
should be non-zero to ensure its vanishing in broken phase.

Thus in the course of the universe expansion and cooling down there existed 
several phases when vacuum energy was non-zero and large by modern
cosmological standards. During inflationary stage the universe was
dominated by vacuum-like energy of the inflaton field. Later on, vacuum 
energy and energy of hot primeval plasma were comparable but still
energy density of matter was dominant, except for periods near possible
first order phase transitions with sufficiently strong super-cooling. 
Vacuum-like energy density should be sub-dominant during big bang 
nucleosynthesis (BBN). Otherwise successful predictions of the latter 
would be distorted. It was possibly sub-dominant during all the time from
BBN to practically present stage and only from redshift 
$z_{eq} \approx 0.67 $ (see sec.~\ref{ss-accel})
the dark energy became dominant and remains such now.

The change of vacuum energy at the electro-weak phase transition is 
about $\Delta \rv^{EW} \sim10^8$ GeV$^4$, which is already huge in 
comparison with $\rho_c \sim 10^{-46}$ GeV$^4$, but at Grand Unification
scale it is by far larger, $\Delta \rv^{GUT} \sim10^{64}$ GeV$^4$.
The change of vacuum energy at QCD phase transition is minor with respect
to those mentioned above, it is ``only'' 45 orders of magnitude larger 
than the cosmological energy density. 

Still, though the QCD contribution to 
vacuum energy is the smallest of all above, it has a very special standing,
because in a sense it is experimentally known quantity. Hadron properties
could be explained in the frameworks of QCD only if there are non-vanishing
quark\cite{quark} and gluon\cite{gluon} condensates. The properties of 
these condensates are well established and it is known that their (vacuum)
energy if at the level $0.1$ GeV$^4$. Thus we know that vacuum is not empty
and the energy density of some particular contributions to vacuum energy
from quarks and gluons is about
45 orders of magnitude larger than the observed value. 

It is difficult to avoid the conclusion that something ``lives'' in
vacuum which very accurately, but not completely, cancels out the energy 
density of the QCD condensates and of some other much larger contributions.
This something is not related to quarks and gluons through QCD interactions
because otherwise it would be observed in direct particle physics
experiments. One may say that very heavy fields/particles may escape
experimental
observations but it is hardly possible that a heavy field may take care
of vacuum energy at the level of $10^{-46}$ GeV$^4$.

\section{Possible ways to solve the problem \label{s-ways}}

As was already mentioned above, there are two different problems:\\
1) Why vacuum energy is not infinitely, or almost infinitely, large?\\
2) What creates the observed universe acceleration? Is it a new form
of energy or gravitational forces are modified at cosmologically large
distances? \\
In what follows we will concentrate on the first problem 
because it seems very likely that its solution should give an
insight into the physical nature of the dark energy. On the other hand,
more observational data on phenomenology of the dark energy
may present an important clue to the solution 
of the first problem.

The simplest, but probably unsatisfactory, suggestion is to say that
the sum of all contributions to vacuum energy is canceled out by a
{\bf subtraction constant}, $\rho_{sub}$, which is chosen quite precisely 
to cancel the present day value of the vacuum energy (why today but not 
at some earlier stage of cosmological evolution?) with 100 orders of 
magnitude precision, but not exactly, leaving behind a small remnant 
$\sim \rho_c$. Though it seems impossible to forbid such a 
point of view formally, it does not look very attractive. 

Another approach to attacking the problem is based on {\bf anthropic 
considerations}. If $\Lambda$ is stochastically distributed variable 
(but why?) then one may estimate with which values of $\Lambda$
the probability of life is the largest. Since density fluctuations
which gave rise to galaxy formation became frozen when vacuum energy
started to dominate the probability of life in the worlds with 
a large $\Lambda$ would be small\cite{antrop}. If the probability
distribution of $\Lambda$ is uniform, then one would expect that life 
is most probable in unverses where vacuum energy started to dominate
at the epoch of galaxy formation. It is consistent with what is 
observed in our universe. On the other hand, some time ago, 
prior to the inflationary idea\cite{guth},
the attempts had been done to invoke
anthropic considerations for a solution of the fundamental problems of 
the Friedman cosmology. 
Now, instead of the anthropic solution, we have much better and testable 
cosmological scenario.

A natural suggestion to invoke a {\bf symmetry} which demands 
almost complete vanishing of vacuum energy did not lead to any progress
up to now. Such a symmetry must include bosons and fermions on equal
footing because both kind of fields possess non-vanishing energy-momentum
tensor coupled to gravity and thus such symmetry should be somehow 
related to supersymmetry, which is known to be badly broken. Moreover,
an exact symmetry in flat space-time may be broken in curved one. 
All that makes the symmetry approach very difficult to realize.

{\bf Infrared instability of massless fields in De Sitter 
background}\cite{infr} may somewhat diminish the original vacuum 
energy by quantum back reaction. Earlier works on this 
mechanism\cite{wood} were criticized in papers\cite{dez}
where was argued that the back reaction is too weak to create a
noticeable effect. Activity in this field is continuing and it is
premature to bury it; for recent development see ref.\cite{brand}.

{\bf Modification of gravitational interaction at large distances}
is attracting more and more attention, motivated by higher dimensional
theories, see e.g.\cite{mdf-grv} or by doubling the number of 
gravitons\cite{kogan}. There are much more papers on the subject and it 
is impossible even to mention them all
in this talk, moreover, they are not directly related to the subject 
because they do not address the problem of cancellation of the huge 
contributions into $\rv$ which is the main interest
here. Older approaches to solve the problem of cosmological constant 
with modified gravity pursued the possibility that in a modified theory
the term in the energy-momentum proportional to $\gmn$ does not gravitate.
If such an idea were realized it would solve the problem because in this
case the vacuum energy density would be unobservable. However, it seems
impossible to achieve that because in the course of a phase transition
the equation of state can change from $p=-\rho$ to e.g. $p=\rho/3$ and an
absence of gravity in the first phase and its presence in the second one
would be incompatible with the demand that massless graviton must interact 
with a conserved source.

{\bf Adjustment mechanism} seems most promising to me and because of that
a separate subsection is devoted to it.

\subsection{Adjustment mechanism}

The idea of adjustment mechanism is quite simple: vacuum energy might
stimulate formation of a condensate of some field, coupled to curvature
of space-time, whose energy density compensates the energy density of 
the source\cite{dolgov82}.
In fact this is a general physical principle which was formulated
centuries ago by Le Chatellier. Adjustment mechanism has potential
to solve both problems: to compensate vacuum energy down to acceptable
value and leave behind a non-compensated remnant of the order of 
$\rho_c (t)$. This property of adjustment mechanism had been 
formulated\cite{dolgov82} long before universe acceleration was 
discovered and the problem of vacuum energy attracted common attention.
However, a serious shortcoming of adjustment mechanism is that, up to 
now, none of the models considered in the literature leads to the
realistic cosmology. Still, despite that, some general features of
adjustment could remain in the future more successful models and thus we
will discuss this mechanism here.

In the first attempt\cite{dolgov82} to resolve 
the vacuum energy problem a massless scalar field $\phi$
non-minimally coupled to gravity was introduced. Its Lagrangian has
the form:
\be
{\cal L}_0 = \partial_\mu \phi \partial_\nu \phi /2 +\xi R \phi^2
\label{l0}
\ee
Equation of motion of this field in De Sitter background with constant
scalar curvature $R$ is unstable if the constant $\xi$ is positive.
It corresponds to tachyonic, negative mass squared, case. As a result,
a homogeneous solution of the equation of motion,
$\phi=\phi(t)$, starts to rise exponentially 
while the energy-momentum tensor 
of $\phi$ remains negligible with respect to the vacuum one. Later, when
$T_{\mu\nu} (\phi)$ becomes comparable to $T_{\mu\nu}^{(vac)}$ its 
back-reaction would slow down the expansion rate from the exponential
to a power law and simultaneously the exponential rise of $\phi$ turns
into $\phi \sim t$. So far so good, but the effective gravitational
coupling in this model $G_N = 1/(m_{Pl}^2 + \xi \phi^2) $ is strongly
decreasing with time which seems to be not the case in real world.

There have been several other attempts to implement this idea with a
scalar field but they all were similarly unsuccessful, for a review
and list of references see e.g. the papers\cite{reviews-1,reviews-2}.
It was argued by S. Weinberg (first paper in ref.\cite{reviews-1}) that
there is no-go theorem which does not allow to implement adjustment
mechanism with a scalar field. As we know, however, many or all no-go 
theorems in quantum field theory have been successfully over-went and 
activity in this field still continues.  

An interesting suggestion was made recently in ref.\cite{mukh},
where the authors proposed to modify kinetic term of the compensating
scalar field as follows, 
${\cal L}_{kin} \sim (\partial \phi)^{2q}/R^{2m(2q-1)}$. Since in this case 
the equation of motion of $\phi$ looks roughly speaking as 
\be 
D^2 \phi + F(R,\phi) = 0
\label{phi-r-eq}
\ee
where the function $F$ vanishes at $R=0$ the solution of this equation has
an equilibrium point at $R=0$. However, the authors of this paper
requested that the power $m$ should be sufficiently
large, $m>3/2$, to ensure stability of the solutions. It can be shown that 
in this case the universe expansion always remains so
fast that the contribution of ordinary matter into cosmological energy 
density would be negligible. An attempt to overcome this problem was
done in ref.\cite{dol-kaw} where the models with 
${\cal L}_{kin} = (\partial \phi)^2 /R^2$ or even with
${\cal L}_{kin} = (\partial \phi)^2 /R |R|$ have been considered. 
It was found that there exist solutions of equations of motion
which indeed lead to cancellation of vacuum energy and the cosmological
expansion changes from exponential (De Sitter) regime to the Friedman
regime which could be very close to that in radiation dominated universe.
Some solutions are even stable with respect to small perturbations but
still detailed features of such cosmology are far from realistic. 
Quantization in this approach remains problematic but, first thing first, 
we need to find a realistic solution to the
classical problem, which still remains 
to be found, and after that may start bothering about quantum effects.

Scalar field is not the only possible candidate for the role of compensating
field, higher spin fields, vector\cite{ad-1}, or tensor\cite{dolgov-s2} 
might also do the job. In such models a condensate of time component of
vector field, $V_t$, or tensor field $S_{tt}$ is developed in De Sitter
space-time. The energy of this condensate cancels its 
creator, the vacuum
energy. Though Lorenz invariance is broken in such models, 
possible effects of its
breaking are not dangerous. The unstable classical mode which compensates
positive vacuum energy appears because of ``wrong'' sign of the corresponding
term in the Lagrangian but fluctuations about this background are well
behaved and non-dangerous\footnote{Similar idea with condensation of 
time derivative of a scalar field was discussed recently in ref.\cite{ghost} 
for possible explanation of cosmic acceleration.}.
Unfortunately the versions of the models with vector and tensor fields
do not lead to realistic cosmologies as well. In particular, a special
version of tensor field condensation considered in ref.\cite{ad-1}
leads to a too strong time variation of the gravitational 
constant\cite{grav-var}.

To summarize, no workable adjustment mechanism is found up to now but 
one should not be too pessimistic - it is not yet proved that such 
mechanism cannot exist. At the moment it seems to be the only approach
which may dynamically solve the problems of huge and small
vacuum/dark energies in one blow.

\section{Conclusion \label{concl}}

A very important psychological consequence of the discovery of 
the accelerated cosmological expansion is that it attracted much
deserved attention to the 
vacuum energy problem. However, the bulk of publications on the
subject deal only with the ``small'' part of the problem, namely, 
what is the origin of the acceleration and neglect the ``large'' part -
what is the mechanism of almost exact cancellation of vacuum energy. 
It is quite probable that this two parts ``large'' and ``small'' are
tightly connected and one cannot be understood without the other.

Some enthusiasm is expressed about a solution of the ``large'' problem 
on the basis of higher dimensional theories. However, no noticeable 
success was achieved on this road and anyhow we live in 4-dimensional
world and the problem should be solved there.

As I have already mentioned, my best choice is an adjustment mechanism
and, though all attempts in this direction still did not reach the
goal, the idea, which follows from adjustment ideology, that 
there exists a new form of cosmological energy with an unusual equation 
of state appeared long before before the accelerated expansion
was discovered. So it seems that more work on this mechanism is
desirable and may even be successful in the nearest future.

%\section{Acknowledgements}
%Place acknowledgements at the end of the text.

%

%
\end{document}